\documentclass[preprint,prb]{revtex4}
\usepackage{graphicx}% Include figure files
\usepackage{dcolumn}% Align table columns on decimal point
\usepackage{ifthen}% setup ifthen
\usepackage{rotating}% setup ifthen
\newcommand{\type}{pdf}

\begin{document}

\title {Current-voltage curves for molecular junctions: the effect of substituents}
\author{Charles W. Bauschlicher, Jr.}
\email{Charles.W.Bauschlicher@nasa.gov}
\author{John W. Lawson}
\email{John.W.Lawson@nasa.gov}
\affiliation{Mail Stop 230-3\\
Center for Advanced Materials and Devices \\
NASA Ames Research Center\\
Moffett Field, CA 94035 }

\begin{abstract}
{\normalsize
  We present current-voltage (I-V) curves for phenylene ethynylene oligomers
between two Au surfaces computed using a Density Functional Theory/Greens Function
approach.  In addition to the parent molecule, two different substituents are considered:
one where all the hydrogens are replaced by chlorines and a second where one H is
replaced by an NO$_2$ group.  In this way, we can study the difference between
electron withdrawing and $\pi$  orbital effects.
For low biases, a reduced current for the derived species is
consistent with a shift of HOMO to lower energy due to the electron withdrawal by
Cl or NO$_2$.  At higher biases, the LUMO becomes important, and the Cl and NO$_2$
substituted species carry more current than the parent because the LUMO is 
stabilized (shifted to lower energy) due to the withdrawal of electrons by the Cl and NO$_2$.  
In these molecules, the C$_2$ bridging units as well as the thiol anchor group
are shown to create bottlenecks to current flow.
}
\end{abstract}

\maketitle
\section{Introduction}
\begin{figure}
\ifthenelse{\equal{\type}{ps}} 
{\includegraphics{f1n.eps}}
{\includegraphics{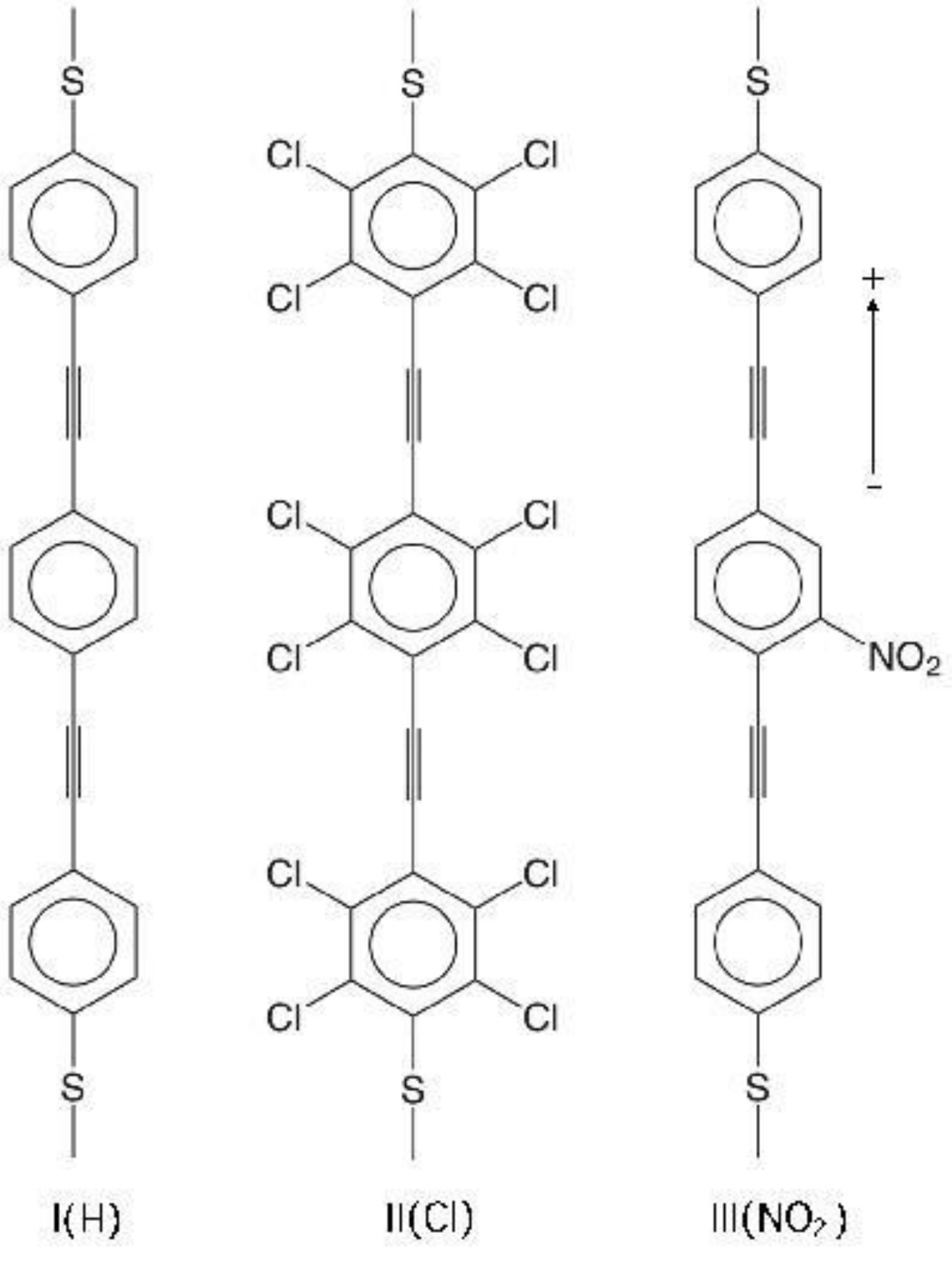}}
\caption{\label{f1} The three molecules studied in this work.  In the free molecule calculations, 
the ends of the molecule are terminated with H atoms, while in the extended molecule, there are six Au atoms bonded to
each of the terminal S atoms. The direction of the
dipole moment for molecule NO$_2$ is shown.  A positive bias voltage is
defined so that the electric field points in the same direction as the dipole moment. }
\end{figure}

    Molecular electronics is a relatively new area, but already many interesting effects
have been observed. One of the more interesting discoveries has been the negative differential
resistance (NDR) found for a phenylene ethynylene trimer, with an NO$_2$ side group, on a gold
surface\cite{chen}, see molecule III in Fig.~\ref{f1}, which we denote as M(NO$_2$).  In
contrast, the parent molecule with all hydrogen atoms (molecule I, which we denote as M(H)) 
does not show NDR.  Since the M(NO$_2$) derivative can have high or low current flow states, it
can be used to store a bit, as Reed and co-workers\cite{reedno2} have demonstrated; they read
and wrote a bit using such a molecular device.\par
    There have been two interesting theoretical investigations\cite{tbs,yin} of related molecules.
Taylor et al.\cite{tbs} studied molecules M(H) and M(NO$_2$) between  two Au surfaces and found 
very similar current-voltage (I-V) curves for these two molecules.  More recently Yin et
al.\cite{yin} studied molecule M(H) and species related to M(NO$_2$).  They found that adding
NO$_2$ to the central benzene ring and an NH$_2$ group to either the central or end benzene rings
increased the current flow relative to the unsubstituted species. They concluded that the conduction
was through the lowest unoccupied orbital (LUMO) and the NO$_2$ shifts the LUMO to lower energy
and hence the current increases relative to M(H).  Since NH$_2$ has little effect on the LUMO,
their results suggest that M(NO$_2$) would carry more current than M(H).  Yin et al. also noted
that the addition of NO$_2$ significantly affected the shape of the highest occupied molecular
orbital (HOMO), but that it is difficult to directly relate the nature of the orbitals to the
I-V curves.  We made the same observation for related molecular systems\cite{cwb2}.  In this
manuscript we compare the I-V curves for the three related molecules shown in Fig.~\ref{f1}.
Molecule II, which we denote as M(Cl),  has not been studied experimentally, but it is studied
here because it is expected to shift the orbital energies relative to M(H), like molecule M(NO$_2$).
However, M(Cl) is not expected 
to affect the character of the $\pi$ orbitals as found for M(NO$_2$).  That is, a 
comparison of M(Cl) and M(NO$_2$) can yield some insight into electron
withdrawing and $\pi$ orbital effects.
\par
\par
\section{Methods}
    The I-V curves are computed using the self-consistent, non-equilibrium, Green's function approach as 
implemented by Xue, Datta, and Ratner\cite{xr1,xr2,xue031,xue032,xueth}.    Our approach has been described in 
detail in previous work and we only summarize it here.  We include six Au atoms from each surface in our treatment
of the extended molecule.   The extended molecule is coupled to two semi-infinite gold (111) surface 
with the 6 Au atoms removed, whose effects are included as self-energy operators through a recursive
Green's function procedure.  The coupling between the bulk contacts and the extended molecule is determined
using a tight-binding approach\cite{xr2,Papa86}, where an additional 27 gold
atoms in each contact are coupled directly to the extended molecule.  Thus the calculations correspond
to a single isolated bridging molecule between two Au(111) surfaces and not to a calculation including periodic boundary
conditions.  \par
	The extended-molecule electronic structure 
calculations are based on density functional theory (DFT), using the pure BPW91\cite{becke,pw91} functional.  
The $\alpha$ and $\beta$ spin densities are constrained to be equal in the extended molecule calculations.
The Au atoms are described using the Los Alamos effective core potential\cite{lanl1} with 11 valence electrons.
As in previous work, the most diffuse s, p, and d primitives are deleted from the associated
valence basis set, and the remaining primitives are contracted to a minimal basis set.  The C, O, N,
Cl, and S atoms are described using the compact effective core potential and the associated 121G basis
set\cite{sbk84}, i.e. the CEP-121G basis set.  A d polarization function is added\cite{popleb} to
the C, O, N, Cl, and S atoms, and diffuse functions are added to O, N, Cl and S. The hydrogen set is
the 6-311G set developed by Pople and co-workers\cite{popleb}.  This valence triple zeta basis set
is the VTZ+P set used in most of our previous work.  We use a temperature of 300~K in the Green's
function calculations.\par 
	We should note that in previous work\cite{cwb3,alle} we considered larger metal clusters and, while
I-V curves obtained using the Au$_6$ clusters are not completely converged with respect to the
size of the metal cluster, they are qualitatively correct.  Because the three molecules considered
in this work are similar and we are interested in relative differences, the use of the Au$_6$
clusters is a good compromise between accuracy and computational expense.
\par
   The bridging species studied are derived from the three molecules shown in Fig.~\ref{f1}.  Their
geometry is optimized at the B3LYP/6-31G* level\cite{hybrid,b3lyp,popleb}.  The terminal H atoms were
removed and the fragment is connected to the two Au(111) surfaces.  The $C_2$ axis of the molecular 
fragment is perpendicular to the surfaces and the S atoms are placed above a three-fold hollow 
at a distance of 1.905~\AA\ above the Au surface.  We should note that a full optimization of
M(NO$_2$) results in a back bone that is not linear, so the optimization is constrained so that the 
$C_2$ axes of the benzene rings and the connecting C$_2$ units are all colinear.  This is
consistent with the other studies.
\par
    We report results for both zero bias transmission functions as well as full I-V
characteristics.  The transmission function is calculated using the Landauer equation
$T(E)=Tr[\Gamma_R G \Gamma_L G^{\dagger}]$ where $\Gamma_R,\Gamma_L$ are the
coupling functions for the right and left
contacts.  The current is evaluated as an integral of $T(E)$ in an energy window around
the Fermi level,  $$ I= \frac{2e}{h} \int_{-\infty}^{\infty} T(E)\times [f(E-\mu_l)-f(E-\mu_r)] dE,
$$ where $f$ is the Fermi function.  The current is of direct interest since it corresponds to an
experimentally observable quantity.  The transmission spectrum, while directly related to the current,
also contains important microscopic information.  \par
	In this work we compute the change in some properties, like the charge density and
electrostatic potential, due to contact formation.  These are computed as the property of the extended molecule 
(connected to the bulk at zero bias) minus the property of the free molecule minus the property
of the two Au$_6$ clusters (connected to the bulk).  \par
	The electronic structure calculations are performed using the Gaussian03 program system\cite{gaussian}.
All of the Green's function calculations are performed using the code described previously
\cite{xr1,xr2,xue031,xue032,xueth} that has been modified for the hybrid and analytic integration\cite{alle}.
\section{Results and Discussion}
     The electron affinities (EAs) and selected orbital energies of the three molecules studied in this work are 
given in Table~\ref{f1}.  We first note that the EA values are M(H)$<M($NO$_2$)$<$M(Cl), and the orbital energies 
are consistent with the EA values.  Namely, the twelve Cl atoms withdraw more electrons from the rings
than the one NO$_2$ group, which withdraws more than the all hydrogen atom case.  This electron withdrawal stabilizes the orbitals of M(Cl) the most, followed by M(NO$_2$) and lastly by M(H).   \par
   We plotted the orbitals of the free molecules and the  extended molecule (i.e. the bridging molecule connected to two Au$_6$ clusters).  Since neither set of orbitals appears to offer great insight into the
conduction, we do not show the plots, but we note the character of the orbitals.  First considering
the free molecule orbitals.
The HOMO and LUMO for molecules M(H) and M(Cl) shows that they are delocalized and are
very similar in character, with the LUMOs being even more similar than the HOMOs.
That is, substituting Cl for H has shifted the orbital energies, but has not significantly
affected the nature  of the HOMO and LUMO.   For molecule M(NO$_2$), Yin et al.\cite{yin}, who found that the HOMO was localized  mostly on the NO$_2$ group and the  LUMO was delocalized, while at
the level of theory used in our work, our HOMO is delocalized, but our LUMO is localized.  Clearly, the localization depends on the
choice of functional used. Using the extended molecule orbitals, 
one finds that the HOMO and HOMO-1 of all three molecules are essentially metal-S sigma bonds.  The HOMO-2 is a $\pi$ orbital
on the bridging molecule, with a sizable component on the metal, and
looks very similar for all three molecules.  The  LUMO is a $\pi$ orbital and  also looks very similar for all three molecules considered.  A notable difference for the three molecules is the LUMO+1 for M(NO$_2$) which
is mostly localized on the NO$_2$ group.\par
\begin{figure}
\ifthenelse{\equal{\type}{ps}} 
{\resizebox{150mm}{!}{\includegraphics{f2n.eps}}}
{\resizebox{150mm}{!}{\includegraphics{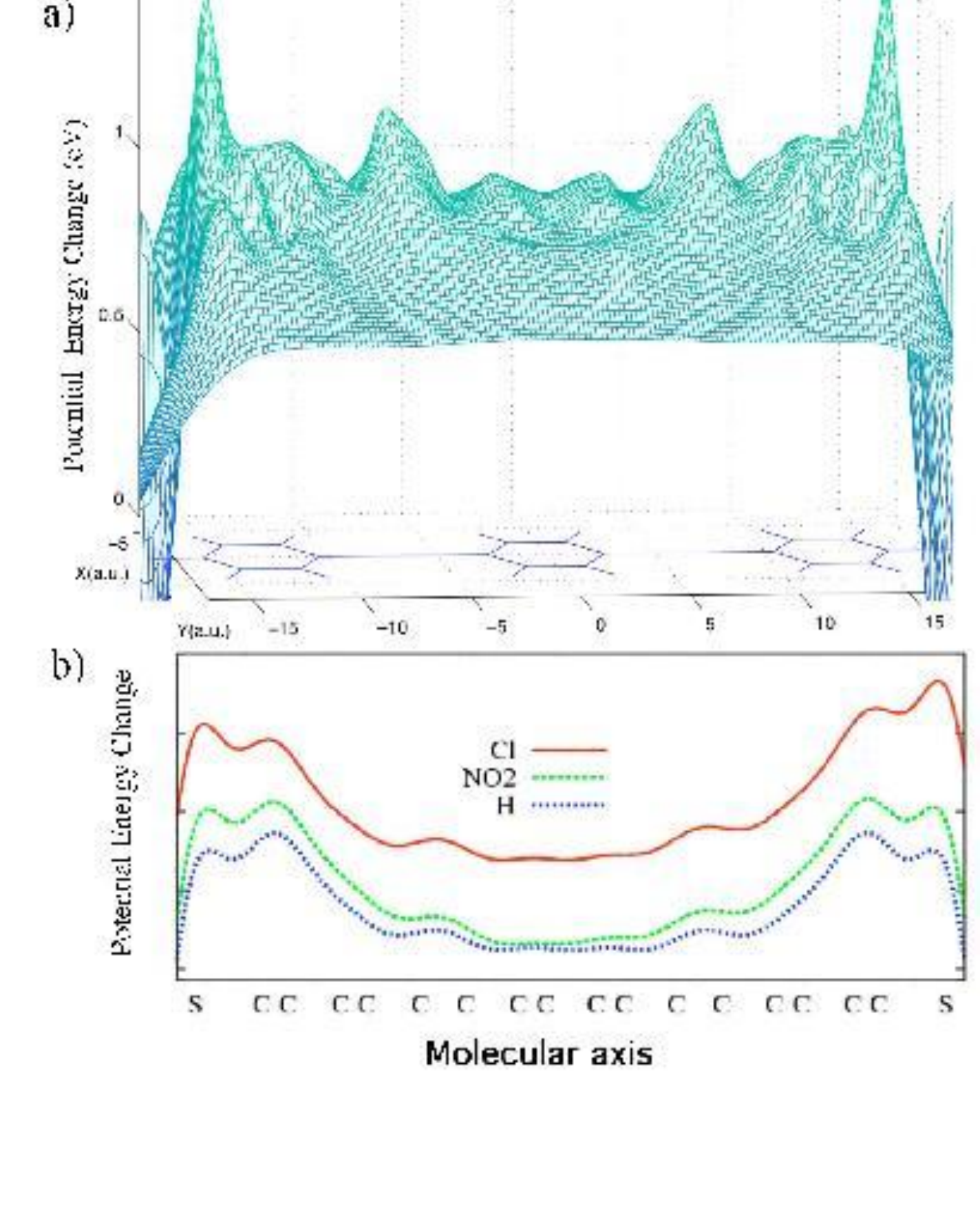}}}
\caption{\label{f2} (Color online) The change in the electrostatic potential energy due to contact formation.
a)  for molecule I, M(H) and b) a comparison of the 3 species considered in this work.}
\end{figure}

   When the molecules are connected to the  bulk and an electric field is applied, the orbitals 
will mix, making it difficult to interpret how the nature of the
molecular orbitals will affect the I-V curves.  To obtain a more accurate picture of the factors  affecting
conduction, we investigate properties computed with the molecule connected to the bulk.  We computed
both the change in charge density and in the electrostatic potential energy due to contact formation. Since the information
obtained from both properties is similar, we plot only the electrostatic potential energy in Fig.~\ref{f2}.
The electrostatic potential energy for molecule M(H) with
the z-coordinate integrated out is shown in Fig.~\ref{f2}a, while in Fig.~\ref{f2}b we compare the change in
electrostatic potential energy for all three molecules along the axis of the molecule.
For M(H)  there are large changes at the ends of the molecule, but there are also sizable changes at 
the C$_2$ bridging units; not surprisingly, there were changes in the charge density
at the same two locations.  It is not too surprising to see large changes were the
Au-S bond forms and even some changes on the benzene ring nearest the S atoms.  However, we find
it somewhat unexpected that the C$_2$ bridging units show larger changes than some of the C atoms in the
end benzene rings.  It appears that forming the Au-S bond has affected the electrostatic
potential (and charge density) throughout the molecule.  Fig.~\ref{f2}b shows that  M(H) and M(NO$_2$) are fairly similar, however, it is perhaps a bit surprising that the
biggest differences are at the ends of the molecule and not in the center where the NO$_2$ is located.
The plot for M(Cl) shows larger differences with M(H) than does M(NO$_2$), which is consistent with 
larger electron withdrawing power of the Cl leading to larger changes for the  M(Cl) density compared with 
M(H) and M(NO$_2$). The barrier heights at the Sulfur atoms reflect the difficulty for electrons to get onto the bridging molecule.
From these plots, we might predict that M(H) would have the higher current at a given voltage, followed by M(NO2),
and finally by M(Cl).  We will see that calculations of the I-V curves bear this out. \par

\begin{figure}
\ifthenelse{\equal{\type}{ps}}
{\includegraphics{f3n.eps}}
{\includegraphics{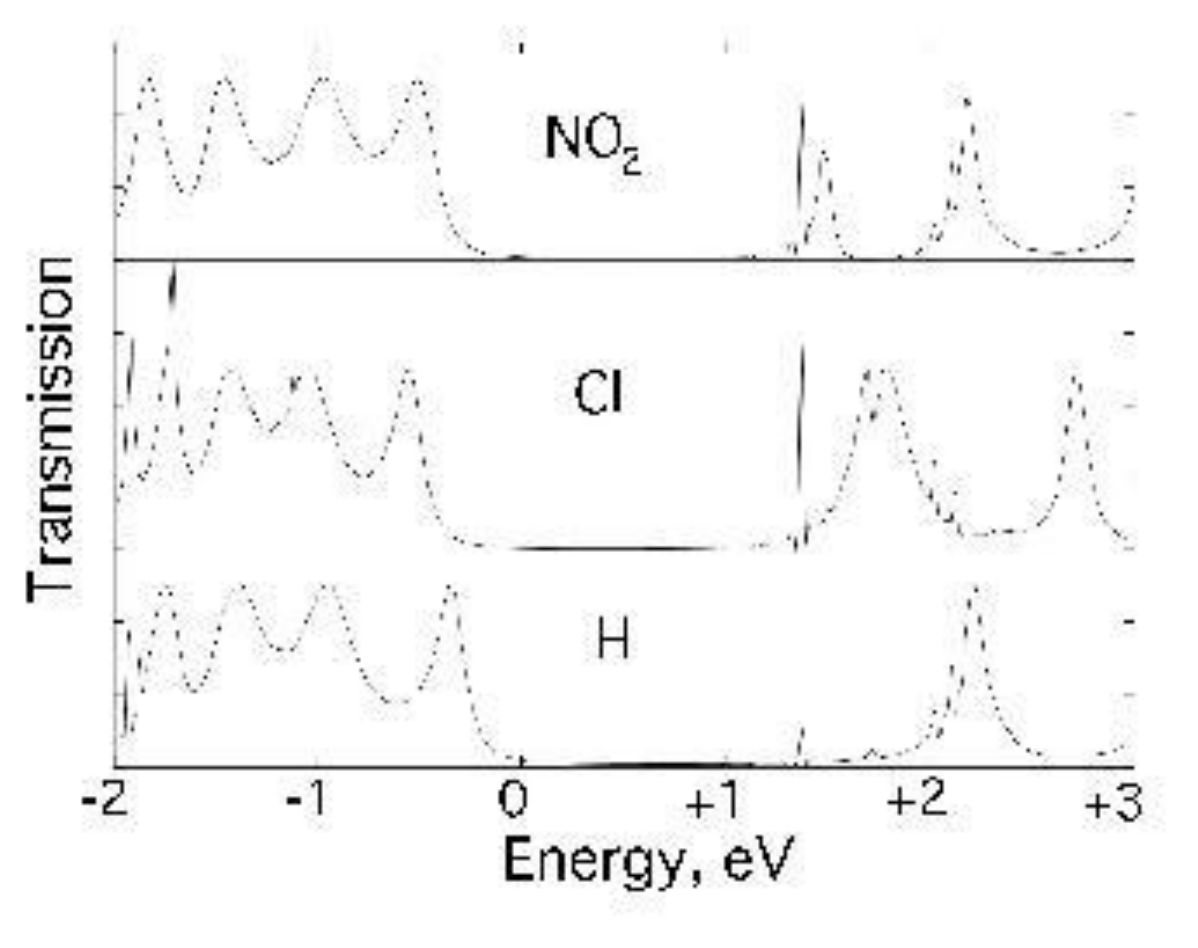}}
\caption{\label{f3} The transmission coefficient for the three molecules
studied in this work.  The Fermi level is set to 0. }
\end{figure}

   The transmission coefficients for the three molecules are plotted in Fig.~\ref{f3}.   The Fermi level has
been shifted to zero.  An inspection of these plots shows that the HOMO lies close to the Fermi level and at
low bias voltage, it will dominate the conduction.  Since the addition of NO$_2$ or Cl shifts the orbitals to
lower energy, the HOMO for these molecules is further from the Fermi level than for the parent molecule M(H).
Therefore, molecules M(Cl) and M(NO$_2$) will have lower conduction than M(H) at low voltages. These electron
withdrawing groups also shift the LUMO closer to the Fermi level, so that at higher biases the conduction for
M(Cl) and M(NO$_2$) should exceed M(H).  The shift to lower energies for molecule M(Cl) is larger than for
molecule M(NO$_2$), therefore M(NO$_2$) will conduct better than M(Cl) at low voltages, but the larger peaks
for the virtual orbitals (1.6-1.7~eV) of M(Cl) suggests that at still higher voltages molecule M(Cl) may
have the highest conduction.  \par
\begin{figure}
\ifthenelse{\equal{\type}{ps}}
{\resizebox{150mm}{!}{\includegraphics{f4n.eps}}}
{\resizebox{150mm}{!}{\includegraphics{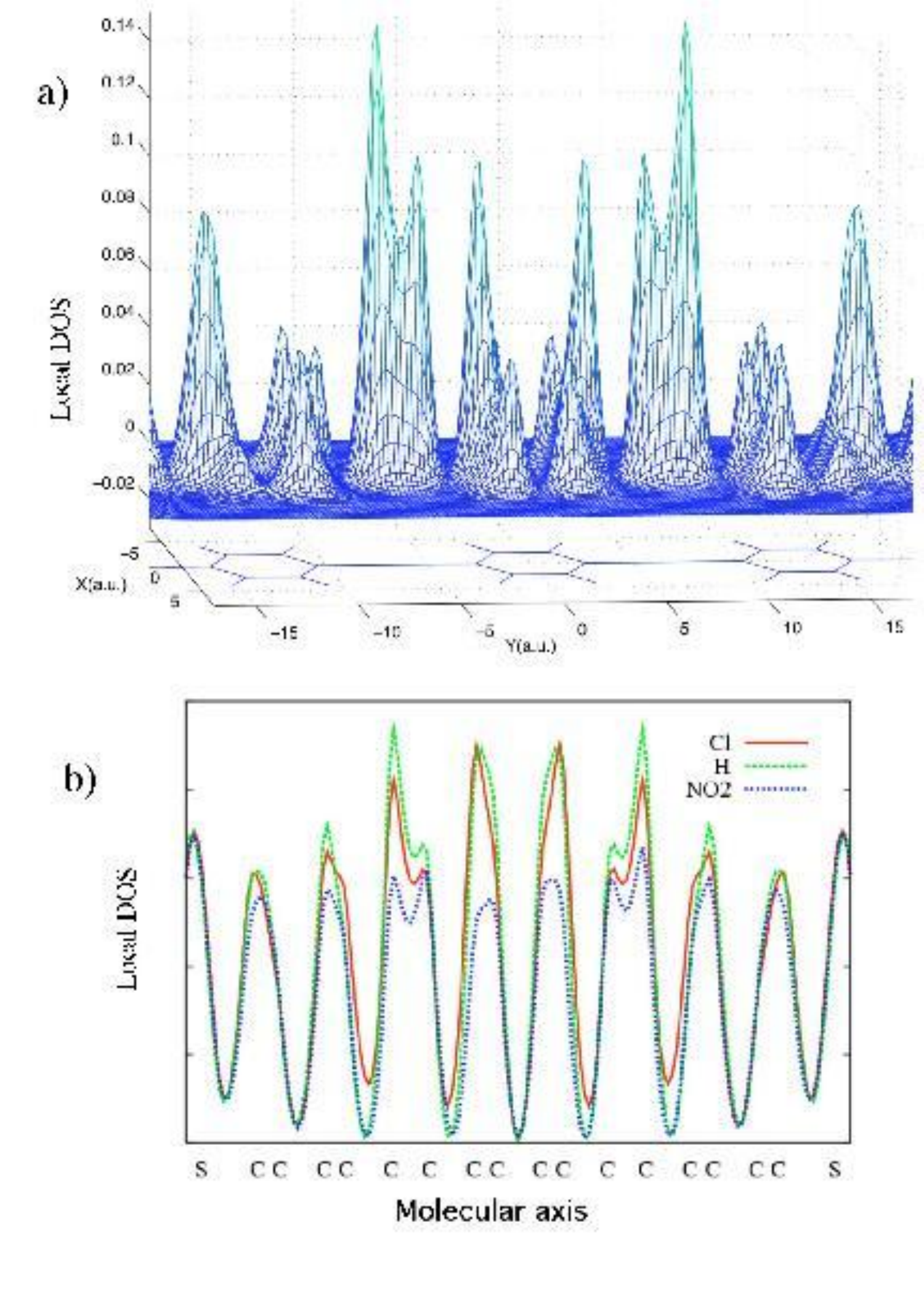}}}
\caption{\label{f4} (Color online) Local density of states for the first
channel below the Fermi level.   a)  for molecule I, M(H) and b)
a comparison of the 3 species considered in this work.}
\end{figure}

   Using the transmission coefficients, it is possible to identify  conduction channels for the
molecules bonded to the metal surfaces.  The local density of states (LDOS) gives a spatial profile of these
channels.  For convenience we refer to the first channels above and below the Fermi level
as the HOMO and LUMO channels, respectively.  Note however that these channels do not correspond
to the HOMO or LUMO orbitals of the parent molecules.
The LDOS  of the HOMO channel of molecule M(H) is plotted in Fig.~\ref{f4}a. It 
looks like the HOMO of the free molecule.  In Fig.~\ref{f4}b we plot the LDOS for the HOMO
channel of all three molecules along the axis of the molecules and
where we have integrated over the x and z directions.  The M(H) and M(Cl) curves are very similar.  The
curve for M(NO$_2$) shows a larger difference with M(H) than does M(Cl).\par
\begin{figure}
\ifthenelse{\equal{\type}{ps}}
{\resizebox{130mm}{!}{\includegraphics{f5n.eps}}}
{\resizebox{130mm}{!}{\includegraphics{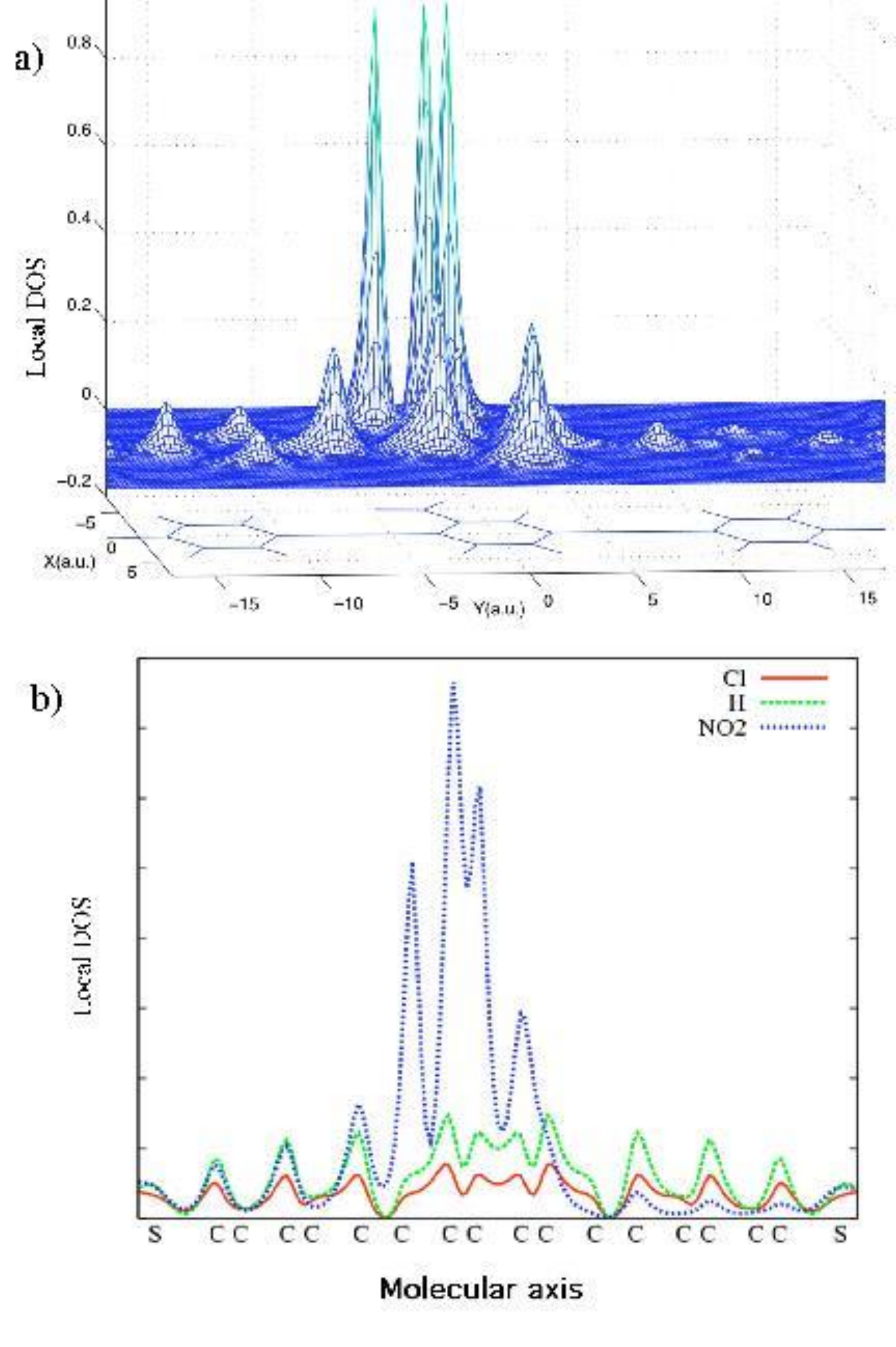}}}
\caption{\label{f5}  (Color online) Local density of states for the first channel 
above the Fermi level. a)  
for  molecule III, M(NO$_2$) and b) a comparison of the 3 species considered in this work. }
\end{figure}

   In Fig.~\ref{f5}a we plot the LDOS for the LUMO channel of M(NO$_2$).  As with the LUMO of the
free molecule, it is localized mostly on the NO$_2$ group and the central benzene ring.  The integrated
local density of states for the LUMO channels of the three molecules are shown in Fig.~\ref{f5}b.  As expect, the M(NO$_2$)
plot is qualitatively different from those for M(H) and M(Cl).  It is interesting to note that the local density of
states associated with the LUMO channel of M(H) and M(Cl) are more different than are their HOMO
channels.  This is the
reverse of the orbital plots for the free molecules where the LUMOs looked more similar than the HOMOs.  
Such changes are to be expected since there are significant changes in the molecule associated with
bonding to the metal.  This is another reminder that while some insight can be obtained from the
orbitals of the free molecules, one  must show caution and not over interpret the free molecule results.  It is 
more reliable to compute the local density of states.\par
\begin{figure}
\ifthenelse{\equal{\type}{ps}}
{\includegraphics{f6n.eps}}
{\includegraphics{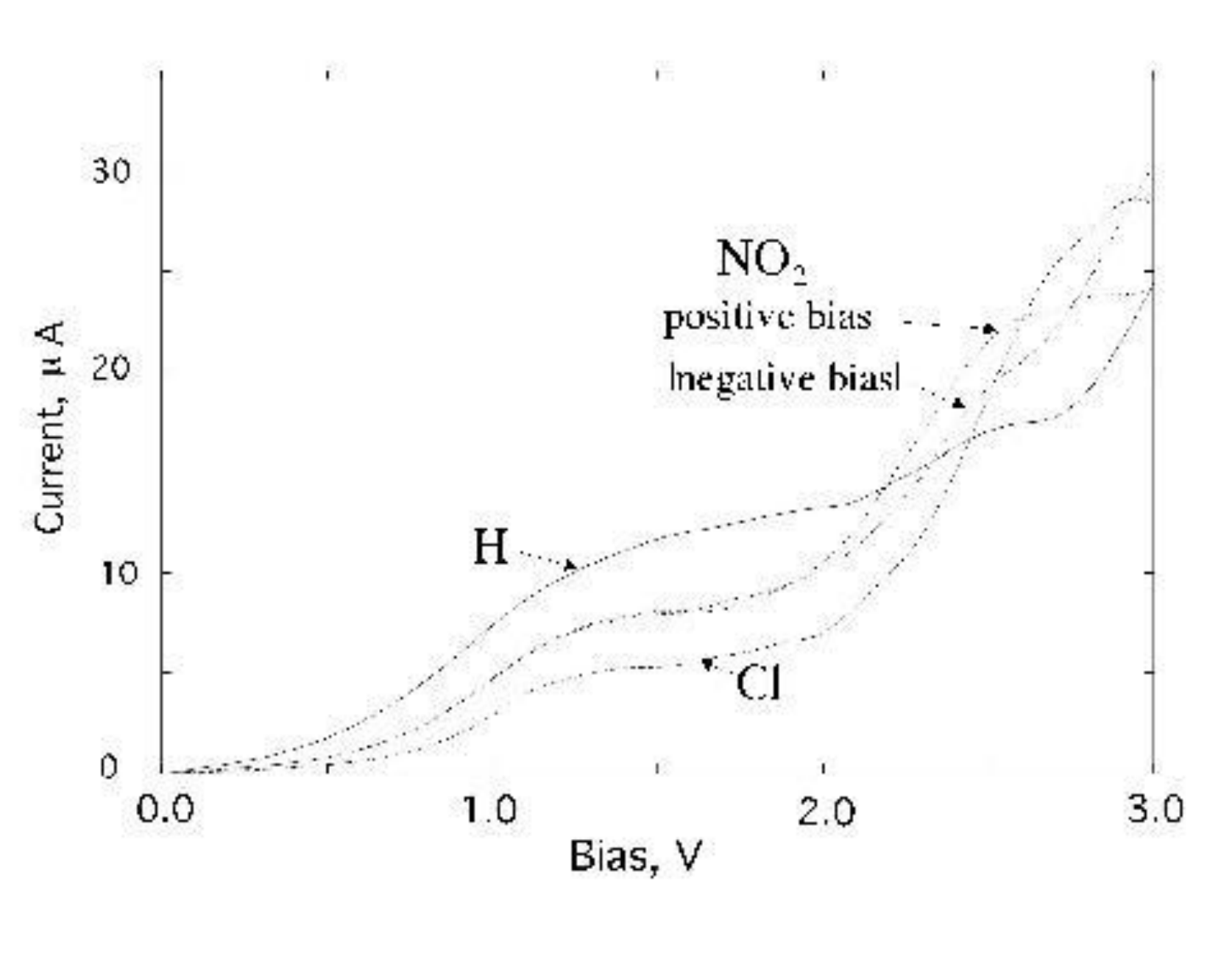}}
\caption{\label{f6} I-V curves for the three species. }
\end{figure}

   The computed I-V curves for all three molecules are shown in Fig.~\ref{f6}. 
Before discussing those computed I-V curves, we note that molecules M(H) and M(Cl) are symmetric, and
therefore their I-V  curves for positive and negative biases are the same.  Molecule M(NO$_2$) is asymmetric and 
therefore its I-V curves for positive and negative biases are different.   Therefore in  Fig.~\ref{f6} we plot
the full I-V curve for molecule M(NO$_2$) with the negative biases plotted as the absolute value 
of the current to more clearly show the small difference between the positive and 
negative bias voltages.  Free M(NO$_2$) has a
small dipole moment along the backbone (1.11~Debye).  While the dipole moment is small, the polarizability
along the backbone is very large (922~$a_0^3$), and therefore at relatively low fields (5$\times$10$^{-4}$~a.u.),
the molecule is stabilized for both a positive and negative field.  That is, at low fields the polarizability
dominates the dipole moment.  Therefore, it is not surprising that there is only a very small difference in 
the I-V curves at low bias voltage.  Yin et al.\cite{yin} also found a small difference between the positive 
and negative biases for similar molecules.   Above 2~V we find a small difference between the positive and
negative biases, and rather unexpectedly a crossing of the I-V curves at about 2.7~V.  Plots of the transmission
coefficients at these biases suggest that at higher voltages (i.e. higher electric fields) the HOMO and
LUMO channels are  affected differently by the positive and negative fields.  If the free molecule is placed in positive
and negative electric fields, the valence orbitals mix.  For example, the HOMO and HOMO-1 mix and localize
one on one side of the molecule and one on the other.  The unoccupied orbitals also mix.  An inspection
of the orbital energies shows that some orbitals are stabilized (or destabilized)  by both a positive and
negative fields, while some are stabilized by one field and not the other. Given all the changes that occur,
it is probably not too surprising that there are some differences in the shape of I-V curves for positive and
negative biases.  We should also note that in the past we have found\cite{cwb1} bumps in the I-V curves
that were related to basis set limitations.  It is possible that some of these differences arise
from limitations in our ability to describe the distortion induced in the orbitals by the larger fields.
However, considering that we are using the valence triple zeta basis set, we suspect that basis set artifacts
should be small.\par
    We now focus on comparing the  I-V curves for all three molecules.  It is fair to say that 
    the computed I-V curves correspond to our expectations based on the
zero bias transmission coefficients.  Namely, molecule M(H) has the largest current at low bias voltages, but
as the bias is increased the values for M(Cl) and M(NO$_2$) increase, eventually surpassing the values for
molecule M(H).  We note that our results differ from previous theoretical results. Taylor et al.~\cite{tbs}
found essentially no difference in the current for M(H) and M(NO$_2$).  While Yin et al.~\cite{yin} did not
consider M(NO$_2$), they considered similar molecules and argued that the conduction was through the LUMO
and hence the reduction in the orbital energies by the NO$_2$ group would increase the current.  We are aware
of an experimental study by Xiao et al.~\cite{xiao} that measured the I-V curves for M(H) and M(NO$_2$) between
0 and 1.5~V.  They found that, in this range of bias values, the current of M(NO$_2$) was half that of M(H).
Our computed results are in good agreement with this.   However, we should note that the total current in
experiment is about two orders of magnitude smaller than that found in our calculations. This is typical for these
types of calculations.  In addition, the experimental results of Xiao et al. found NDR at higher voltages for molecule
M(NO$_2$), which we do not see in our calculations.\par
\begin{figure}
\ifthenelse{\equal{\type}{ps}}
{\resizebox{150mm}{!}{\includegraphics{f7n.eps}}}
{\resizebox{150mm}{!}{\includegraphics{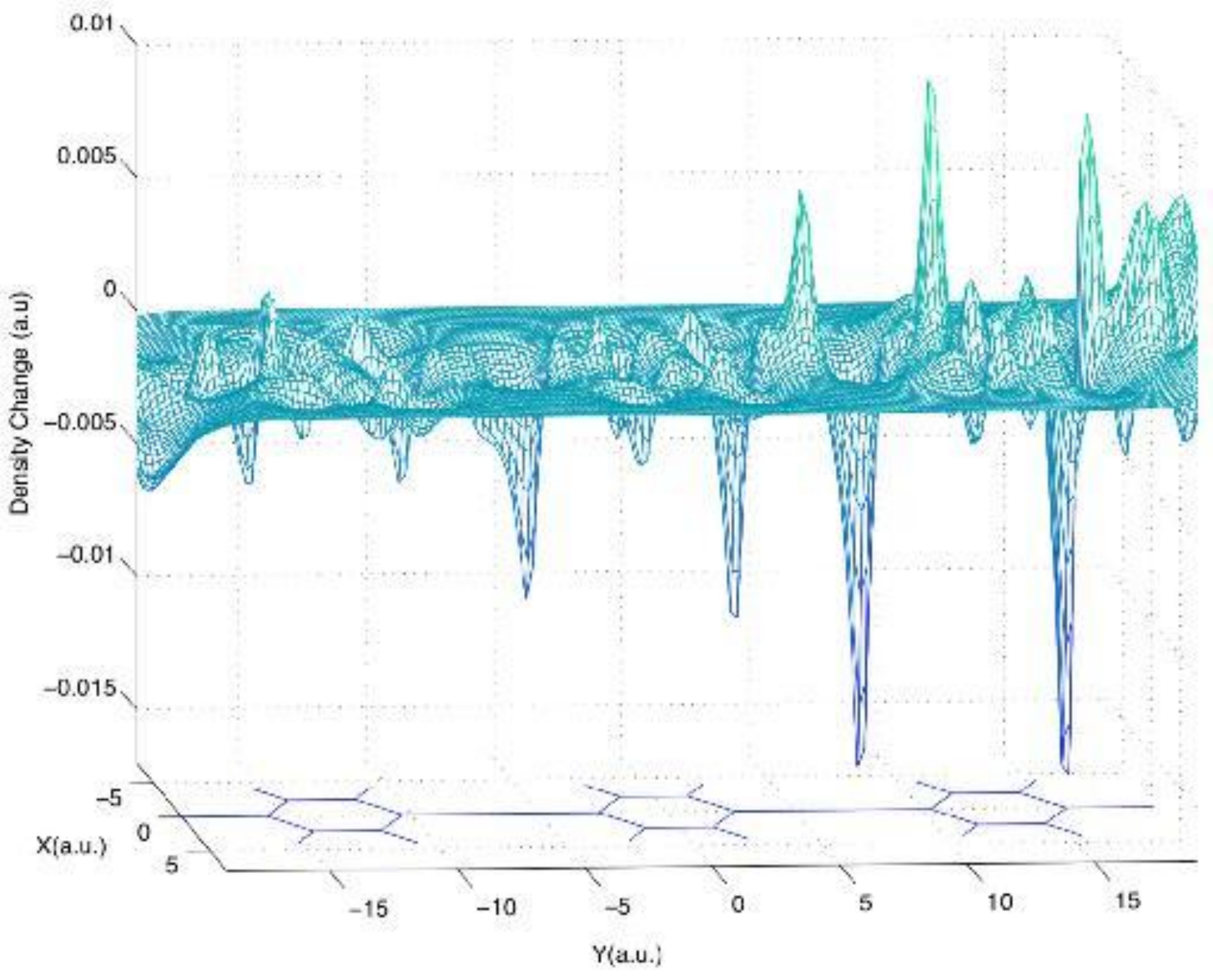}}}
\caption{\label{f7} (Color online) Change in the charge density for molecule I, M(H), for an
applied bias of 2.0~V relative to equilibrium.}
\end{figure}

    In Fig.~\ref{f7} we plot change in charge density at an applied bias of 2~V relative to equilibrium (i.e.
no bias).  The build up of charge at one end of the molecule is consistent with similar plots for benzene-1,4-dithiol,
where only one benzene ring between two Au surface was considered \cite{xue031}.  In that previous work, 
charge build up near the S atoms resulted in ``resistivity dipoles" that impeded current flow. 
In the molecules we considered, there is, in addition, a significant build up of charge
at the C$_2$ bridging units, showing that they also act as a bottleneck to charge flow.   Perhaps
these additional C$_2$ bottlenecks to charge flow help explain why there is a significant difference between 
the I-V curves for M(H) and M(Cl).  In previous work, replacing the H atoms with Cl for a benzene-1,4-dithiol molecule,
which has no C$_2$ bridging units, has only a  little affect on the I-V curves.  \par
\par
\section{Conclusions}
    We find conduction at low bias values is through the HOMO  for the molecules we considered.
Therefore the substitution of Cl atoms or an NO$_2$ group for the hydrogen atoms of molecule M(H) stabilizes the HOMO and
reduces the current of the substituted species relative to the parent.  This reduction in current for the NO$_2$
species is consistent with experiment.  However, the computed current is about two orders of magnitude larger
than that found in experiment, as is typical for even the highest levels of theory.  Analysis of the
results shows that both the C$_2$ bridging units and the thiol anchor groups act as bottlenecks to current flow.  \par
\par
\section{Acknowledgments}
 C.W.B is a civil servant in the Space Technology Division (Mail Stop 230-3), while
J.W.L. is a civil servant in the TI Division (Mail Stop 269-2).\par

\eject
\begin{table}
\caption{\label{t1} Summary of the electron affinity and orbital energies
(in eV) for the three molecules studied in this  work, computed at the BPW91 level of
theory.}
\begin{center}
\begin{tabular}{lrrrrrr}
&\ \ \ \  \ \ \ \          &     M(H)&\ \ \ \ \ \ \ \ & M(NO$_2$) &\ \ \ \ \ \ \ &  M(Cl)   \\
EA&&  1.26&& 1.63&& 2.13\\
\noalign{\vskip 10pt}
Orbital Energies\\
HOMO$-$4&&  $-$6.487&&$-$6.422&&  $-$6.350\\
HOMO$-$3&&  $-$5.912&&$-$6.177&&  $-$6.349\\
HOMO$-$2&&  $-$5.791&&$-$5.824&&  $-$6.231\\
HOMO$-$1&&  $-$5.791&&$-$5.480&&  $-$5.997\\
HOMO  &&     $-$5.100&&$-$4.973&&  $-$5.533\\
LUMO& &      $-$2.637&&$-$3.324&&  $-$3.401\\
LUMO+1&&     $-$1.787&&$-$2.563&&  $-$2.470\\
LUMO+2&&     $-$1.186&&$-$1.672&&  $-$2.151\\
LUMO+3&&     $-$1.186&&$-$1.265&&  $-$2.150\\
LUMO+4&&     $-$1.157&&$-$1.258&&  $-$2.090\\
\end{tabular}
\end{center}

\noindent
\end{table}                                                                                                                         
%equilibrium figures

%non-equilibrium figures

%non-equilibrium figures

\end {document}